# The missing atom as a source of carbon magnetism

Miguel M. Ugeda[1], Iván Brihuega[1]*, Francisco Guinea[2] and José M. Gómez-Rodríguez[1].

[1] Dept Física de la Materia Condensada,Universidad Autónoma de Madrid, E-28049 Madrid, Spain
[2] Instituto de Ciencia de Materiales de Madrid, CSIC, Cantoblanco E-28049 Madrid, Spain
*To whom correspondence should be addressed:  ivan.brihuega@uam.es

**Atomic vacancies have a strong impact in the mechanical, electronic and magnetic properties of graphene-like materials. By artificially generating isolated vacancies on a graphite surface and measuring their local density of states on the atomic scale, we have shown how single vacancies modify the electronic properties of this graphene-like system. Our scanning tunneling microscopy experiments, complemented by tight binding calculations, reveal the presence of a sharp electronic resonance at the Fermi energy around each single graphite vacancy, which can be associated with the formation of local magnetic moments and implies a dramatic reduction of the charge carriers' mobility. While vacancies in single layer graphene naturally lead to magnetic couplings of arbitrary sign, our results show the possibility of inducing a macroscopic ferrimagnetic state in multilayered graphene samples just by randomly removing single C atoms.**

PACS numbers: 73.22.Pr, 75.70.Rf, 73.20.Hb, 68.37.Ef

Graphite is a semimetal, where the low density of states at the Fermi level, and its high anisotropy induces significant differences from conventional metals (1). It consists of weakly van der Waals coupled graphene layers and thus it shows a strong 2D character sharing many properties with graphene. Graphite´s unusual features are enhanced in single layer graphene (2,3), where the density of states vanishes at the neutrality level, and carriers show a linear, massless, dispersion in its vicinity. In graphene, the existence of localized electronic states at zigzag edges (4,5) and vacancies (6) leads to an extreme enhancement of the spin polarizability,  and model calculations suggest that magnetic moments will form in the vicinity of these defects (7-9). It seems likely that similar phenomena also take place in other $sp^2$ bonded carbon materials, such as graphite. An enhanced density of states has been observed near zigzag steps in graphite surfaces (10). The existence of these localized states suggests that magnetic moments (11-13), and possibly magnetic ordering may exist in single layer graphene and graphite. In the case of irradiated graphite, where lattice defects are expected to exist, magnetic order has been reported even at room temperature (14,15).

Introducing vacancies in graphene-like systems by irradiation has been shown to be an efficient method to artificially modify their properties (14-17). While the role played by these vacancies as single entities has been extensively addressed by theory (6-8,18), experimental data available (14-17) refer to statistical properties of the whole heterogeneous collection of vacancies generated in the irradiation process (19,20). The main goal of the present work is to overcome this limitation; thus, we first create perfectly characterized single vacancies on a graphite surface by Ar+ ion irradiation and then, using low temperature scanning tunneling microscopy (LT-STM), we individually investigate the impact of each of such vacancies in the electronic and magnetic properties of this graphene-like system. We identify well localized electronic resonances at the Fermi energy around graphite single lattice vacancies. The existence of these states is in good agreement with theoretical expectations, and it can be associated with the formation of magnetic moments in this all-carbon material. Using simple extensions of these models, as well as the similarities between the properties of a clean graphite surface and single layer and multilayered graphene, we can extrapolate our results to those systems. In addition, we also show that contrary to the single layer graphene case, ferrimagnetism is favored in multilayered graphene samples.

We use highly ordered pyrolytic graphite (HOPG) samples, which present the AB Bernal stacking. Thus, one atom of the honeycomb unit cell (α) is located directly above a C atom of the second layer and the other one (β) is on top of a hollow site (see Fig. 1B). A key point of the present work is the atomistic control of the samples, which was obtained by performing all the preparation procedures and measurements under UHV conditions. We created single vacancies by irradiating with 140 eV Ar+ ions previously in-situ exfoliated HOPG surfaces. At these low ion energies, just above the threshold value for the displacement of surface atoms, the ion irradiation mainly produces atomic point defects (19,20). After further sample annealing at 650ºC, the remaining defects were mostly single vacancies as revealed by STM images. Fig. 1A shows the general morphology of our samples after the irradiation and annealing procedure. The previous perfect and pristine graphite surface, now presents several point defects surrounded by threefold (√3x√3) patterns, R3 in the following (see Fig.S1 in EPAPS (21) for a more general overview). The comparison of our atomically resolved STM images of these defects (Fig. 1A, C) with calculations (22, 23) shows that these defects correspond to single vacancies on both α and β sites of the graphite honeycomb lattice.

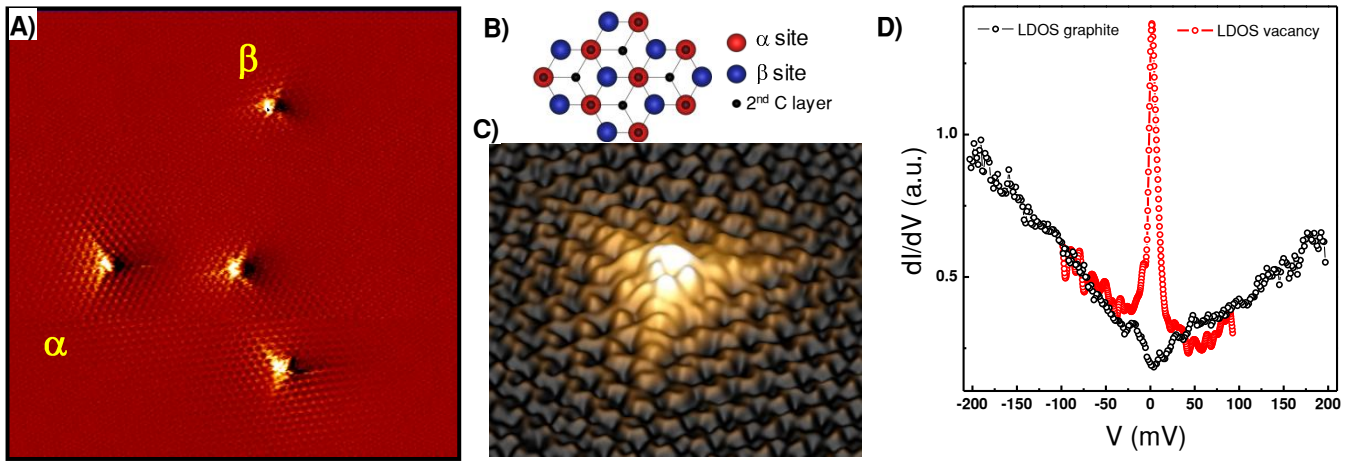

**Fig. 1.**. A) 17x17 nm$^2$ STM topography, measured at 6 K, showing the graphite surface after the Ar$^+$ ion irradiation (for a larger scale overview of the same region see (21)). Data analyzed using WSXM (37). Single vacancies occupy both α and β sites of the graphite honeycomb lattice. Sample bias: + 270 mV, tunneling current: 1 nA. B) Schematic diagram of the graphite structure. C) 3D view of a single isolated vacancy. Sample bias: +150 mV, tunneling current: 0.5 nA. D) STS measurements of the LDOS induced by the single vacancy and of graphite. Black circles correspond to $dI/dV$ spectra measured on pristine graphite and red circles correspond to $dI/dV$ spectra measured on top of the single vacancy, showing the appearance of a sharp resonance at $E_F$. $dI/dV$ measurements were done consecutively at 6 K with the same microscopic tip.

We use a home-made LT-STM (24) to investigate the local electronic structure of the single atomic vacancies created in graphite. This is an unrivaled technique to provide local information about the surface electronic properties, achieving atomic precision and very high energy resolution (≈ 1 meV at 4.2 K). The use of these unique capabilities in graphene-like systems has already allowed, for example, to detect the coexistence of both massless and massive Dirac Fermions in a graphite surface (25, 26), or to prove the Dirac nature of the quasiparticles in epitaxial graphene on SiC (27, 28). Differential conductance ($dI/dV$) spectra were measured in open feedback loop mode using the lock-In technique with frequency 2.3 kHz and a.c modulation of 1 mV. Various tungsten (W) tips were used for the measurements. In order to avoid tip artifacts, tip status was always checked by measuring reference spectra on pristine graphite; only tips showing the standard featureless V-shaped spectra and a work function of 4-5 eV were considered in this work. Spectra remained unchanged for moderate tip-sample distance variations (stabilization current was routinely modified from 10 pA to 10 nA). Fig. 1D shows consecutive $dI/dV$ spectra, measured at 6 K, summarizing our results. Far enough from any defect, spectra showed a featureless V-shaped form as expected for the LDOS of pristine graphite (black circles). Spectra acquired on top of a single vacancy (red circles), both in α and β positions, reveal the existence of a sharp resonance peak around the Fermi level ($E_F$) with a FWHM of ≈ 5 mV.

The presence of this resonance is a fundamental result that, although anticipated in many theoretical works, had never been experimentally observed before. The formation of a magnetic moment can be associated to the resonance, since electron-electron interactions, and the fact that the localized level is very close to the Fermi energy, favour the polarization of this state. In addition, the narrowness of the resonance and the low electronic density of graphite at the Fermi level imply a very poor screening of the magnetic moment, which anticipates a very high Curie temperature for the vacancies (see below). Our results also demonstrate that the presence of these single vacancies should have a strong impact on the electronic transport, since the existence of a resonance in the vicinity of the Fermi level gives rise to a strong reduction of the mobility with a mean free path which tends to the distance between impurities at the neutrality point (29). In this way, the artificial introduction of a chosen density of vacancies can be used as an effective method to tune the mobility of graphite and graphene-like samples.

The existence of this sharp resonance at the neutrality point in single layer graphene can be derived from the nearest neighbor tight-binding Hamiltonian which describes the π bands (6), neglecting deformations near the vacancy. We have checked that the resonance is also present in a semiinfinite graphite layer by extending the calculation and using the Slonczewski-Weiss-McClure parametrization of the π bands, which is expected to describe well the electronic structure near the Fermi level (30-32). Results are shown in Fig. 3B. Details of the calculation are given in the supporting material (21). The results are consistent with those in ref. 6), and with extensions of that model to multilayered graphene (33). A sharp resonance exists when the vacancy is at α sites, while a lower peak is found near vacancies at β sites. The wavefunction Ψ associated to this resonance is very extended, decaying as function of the distance to the vacancy as $r^{-1}$ (6, 33). It shows a R3 modulation, associated to the wavevector which spans the two valleys in the Brillouin Zone (21).

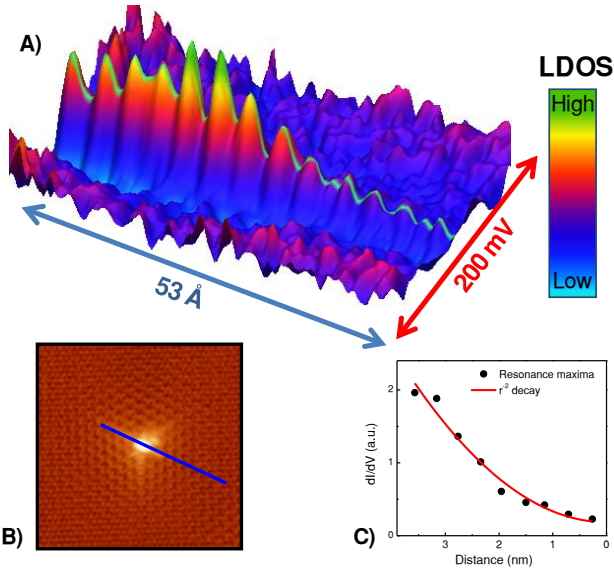

**Fig. 2.** A) LDOS as a function of sample voltage V and position x along the blue line drawn in B). A green line has been drawn to outline the evolution of the resonance peak height, showing a clear R3 modulation. B) STM topographic image of a single graphite vacancy. Image size 8 x 8 nm²; sample bias +200 mV tunneling current 0.6 nA. C) $r^{-2}$ decay of the resonance intensity. Black dots correspond to the maxima of the resonance peak height and the red line is parabolic fit to the experimental data.

We have also analyzed experimentally the spatial extension of the states induced around single vacancies by mapping the narrow resonance as a function of distance from the defect. Fig. 2A shows a map of the local density of states (LDOS) *vs* energy, measured along the line across the vacancy drawn in Fig. 2B. The narrow resonance extends several nanometers away from the vacancy, indicating that it is indeed a quasilocalized state. The resonance shows an overall decreasing intensity with increasing distance, consistent with the expected $r^{-1}$ decay (STM probes $|\Psi|^2$) as shown in Fig. 2C, its height is modulated with the R3 periodicity (Fig. 2A) and its width remains approximately constant for all distances.

The agreement between the experimental results and the theoretical model shows that the latter describes correctly the main electronic properties of the vacancy. Electron-electron interactions prevent double occupancy of the resonance, and lead to the formation of a magnetic moment near the vacancy. The resonance is built up from π orbitals, and it is orthogonal to the extended states. Hence, the coupling between the magnetic moment and the conduction and valence bands is not due to virtual transitions involving short lived zero or doubly occupied states. These are the processes which describe the antiferromagnetic Kondo coupling induced by a magnetic impurity hybridized with a metallic band. In our case, we expect a ferromagnetic coupling mediated by the Coulomb repulsion (21). Moreover, the magnetic moment is extended throughout many lattice cells around the vacancy, so that it interacts with many partial wave channels built up from the extended states, leading to a multichannel ferromagnetic Kondo system (34). The magnetic moment is not quenched at low temperatures.

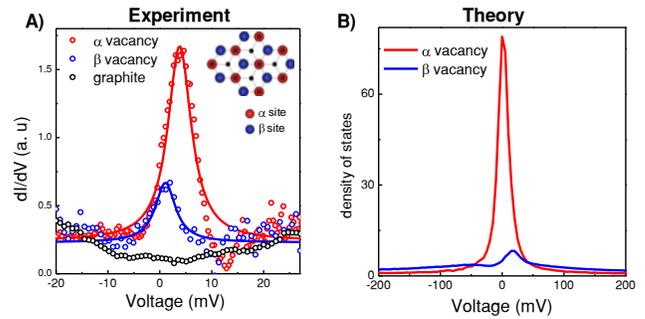

**Fig. 3.** A) $dI/dV$ spectra measured with the same tip on top of a single vacancy in an α (red), β (blue) site and on pristine graphite (black). The intensity of the resonance measured on top of the α vacancy is much higher than the one of the β one, indicating that removing a C atom from an α site generates a stronger magnetic moment. Solid lines are fits to a Lorentzian function giving a FWHM of ≈5mV for both the resonance on the α and β site. Small variations (of a couple of mV) in the position of the resonance peak maxima were observed, which we attribute to the local environment of each specific vacancy. B) Calculated density of states in the atom nearest to vacancy. Red: vacancy in α site Blue: Vacancy in β site.

The interaction between magnetic moments induced by vacancies at different sites has been extensively studied for single layer graphene (35). Its sign depends on the sublattice occupied by the vacancies. In graphite, the two sublattices are inequivalent. Our samples present vacancies in both α and β sites of the honeycomb lattice, which gives rise to two R3 scattering patterns of different shape and extension (21). It is then natural to think that the quasilocalized resonance, and thus the magnetic moment, induced by graphite single vacancies are also affected by the underlying C layers. Our $dI/dV$ spectra clearly demonstrate that this is indeed the case. Fig. 3A shows consecutive spectra measured, with exactly the same microscopic STM tip, on both types of vacancies and on clean graphite. A sharp resonance of very similar width (FWHM of ≈ 5 mV) is induced by both types of vacancies; however, the intensity is much higher in the case of the α vacancy, in agreement with our calculations (Fig. 3B). This inequivalence in the magnetic moment induced by each type of vacancy will reduce antiferromagnetic coupling, inhibiting complete frustration. Hence, we expect a ferrimagnetic ground state at low temperatures. The fact that the resonances form a narrow band of delocalized states suppresses screening effects and fluctuations due to spin waves (36). A simple estimate based on the direct exchange coupling between moments localized around impurities gives a Curie temperature $T_c \approx e^2 \cdot \sqrt{n_v}$, where $e^2$ is the electric charge (≈1-5 eV·Å), and $n_v$ is the vacancy concentration. In the present experiment, $n_v \approx 3 \cdot 10^{11}$ cm⁻² and this simple estimation suggest a Curie temperature $T_c \approx 50\text{-}200$ K (21).

Our findings have strong implications both from an applied and a fundamental point of view. They provide a significant stimulus to the theoretical community demonstrating that for atomistically controlled experiments, tight-binding methods give an excellent description of graphene-like systems physics. The observed resonances indicate that vacancies should limit significantly the mobility of carriers in graphene, and enhance its chemical reactivity. The existence of sharp electronic resonances at the Fermi energy, strongly suggests the formation of magnetic moments around single vacancies in graphite surfaces, implying a magnetic phase for this free of impurities carbon system with high Curie temperatures and small magnetization moments, which indicates a suitable route to the creation of non-metallic, cheaper, lighter, and bio-compatible magnets

We are thankful to J.Y Veuillen and P. Mallet for providing us with the sputtering parameters. I.B was supported by a Ramón y Cajal project of the Spanish MEC. M.M.U. acknowledges financial support from MEC under FPU Grant Nº. AP-2004-1896. Financial support from Spain's MEC under grants No. MAT2007-60686, FIS2008-00124 and CONSOLIDER CSD2007-00010, and by the Comunidad de Madrid, through CITECNOMIK, is gratefully acknowledged.

# SUPPLEMENTARY INFORMATION

## Sample oveview

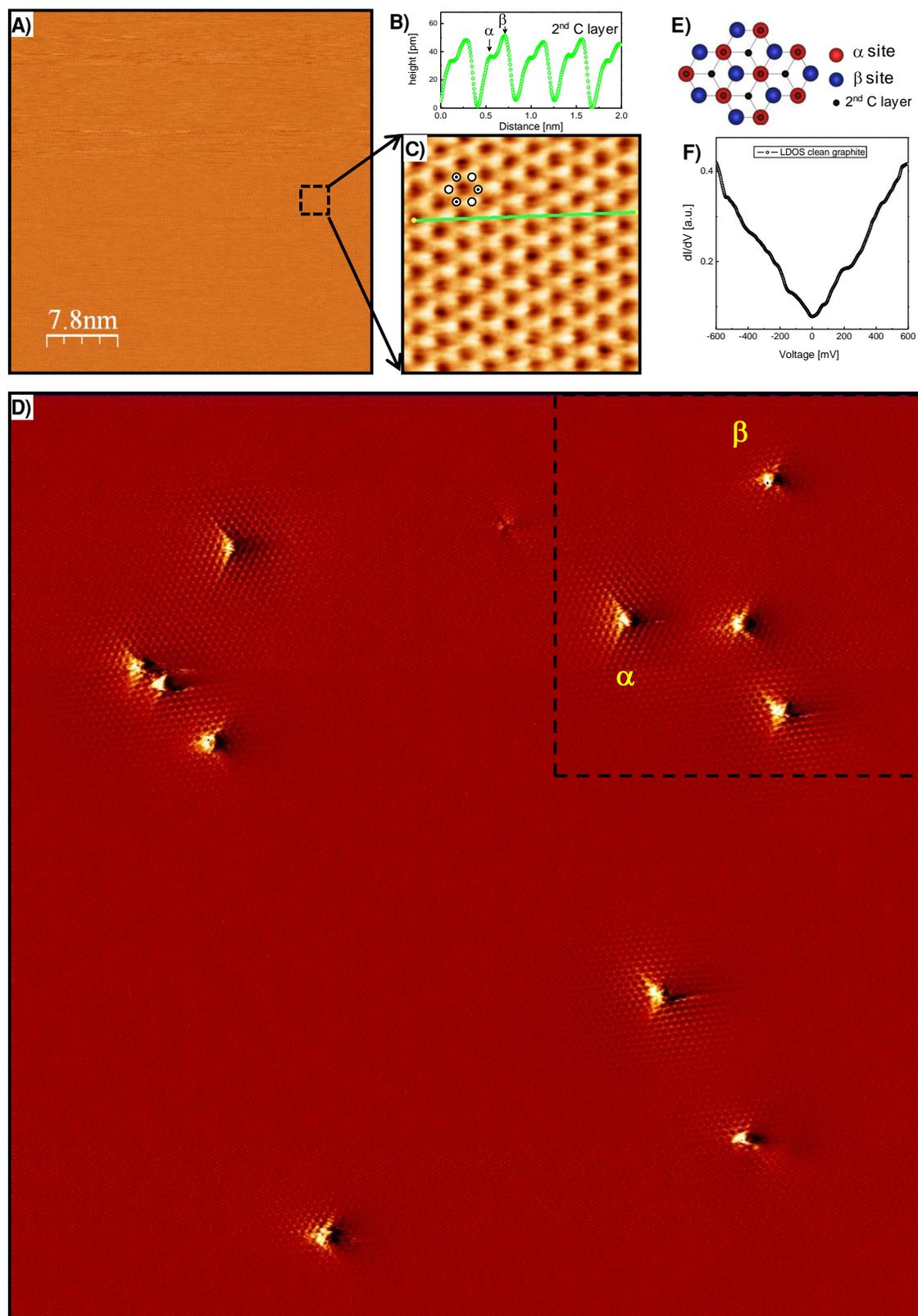

**Fig S1.** Sample overview before and after the irradiation process.

Figure S1 shows an overview of the sample before (a) and after (d) the Ar⁺ irradiation process. In these 40x40nm² images, it can be seen that no defect is present in our surfaces previously to the irradiation procedure. Fig. S1C shows a high resolution STM images of the HOPG surface where both atoms of the honeycomb lattice are resolved; the one on β site showing a higher intensity than the one on α site (see profile in S1B).

The exact atomic location of the vacancies in the graphite lattice can be inferred from STM measurements, as it is reflected in the complex R3 scattering patterns originated from them [22, 23]. Both α and β vacancies exhibit a 3-fold scattering with the presence of three "arms" at 120º each. In the case of Fig. S1D, vacancies on a β site have one of their arms pointing towards left, while in the case of α vacancies one of their arms points towards right.

LDOS of pristine graphite surfaces showed a featureless V-shaped form as inferred from our *dI/dV* spectra (fig S1F).

## THEORETICAL MODEL.

### 1. Electronic structure.

We describe the π bands of graphite using the parametrization suggested in refs. (30, 32), as modified because of recent results for bilayer graphene in ref. (28). The tight binding parameters include hoppings between next and next nearest neighbor layers. The values used (in meV) are shown in Table I.

| | |
|---|---|
| $\gamma_0$ | 3160 |
| $\gamma_1$ | 390 meV |
| $\gamma_2$ | -20 meV |
| $\gamma_3$ | 315 meV |
| $\gamma_4$ | 44 meV |
| $\gamma_5$ | 38 meV |
| $\Delta$ | -8 meV |

The calculation of the local density of states is done used the iterative procedure in (S1). The implementation of the method requires a finite imaginary part of the energy. The vacancy is modeled with a large on site energy, $\varepsilon_{vac} = 2000$ eV. The large enhancement of the density of states near the impurity implies that even the smallest broadenings used, $\approx 1$ meV induce the peak shown in Fig. 3B. A simpler model, which can Table I be solved analytically, is obtained by keeping only the in-plane hopping $\gamma_0$, and nearest neighbor interlayer hopping, $\gamma_1$. The analytical solution of this model gives, for vacancies in an α site the same solution as that found for single layer graphene (6, 32), localized in the layer where the vacancy sits. The density of states at bulk and surface states are shown in Fig. S2.

The resonance in single layer graphene (6) is very delocalized, $|\Psi(r,\theta)|^2 \approx 1/(r^2 \log(D/a))$ where $a$ is the lattice constant, and $D$ is the distance between vacancies. It can be smoothly matched to a superposition of continuum solutions obtained from the Dirac equation. Because of that, a local perturbation which induces a change near the position of the vacancy does not affect much the resonance. A perturbation of strength $V_0$ localized within a length $d$ of the vacancy changes the position of the resonance by $\Delta \varepsilon \approx V_0 \int_0^d r |\Psi(r)|^2 \approx V_0 \log(d/a)/\log(D/a)$. Hence, for $d \approx a$ and $D >> a$, the shift in the resonance away from the neutrality point is small. As the extended states have a density of states $D(\Delta \varepsilon) \approx \Delta \varepsilon / v_F^2$, the broadening of the resonance is also small.

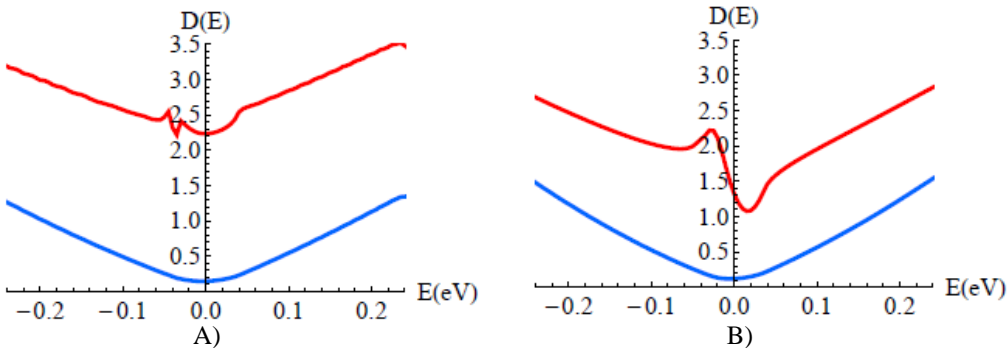

**Fig, S2**. A) Density of states at a bulk site. Blue: α site. Red: β site. B) As in A), for a surface site.

## 2. Interaction effects.

The interactions are described by the Hamiltonian:

$$H = H_0 + H_{\text{int}} \equiv t \sum_{i,j;s} c_{i;s}^+ c_{j;s} + \sum_{i;s} \varepsilon_i c_{i;s}^+ c_{i;s} +$$

$$+ U \sum_i \left( n_{i;\uparrow} - \frac{1}{2} \right)\left( n_{i;\downarrow} - \frac{1}{2} \right) + \sum_i V(\vec{r}_{ij})\left( n_i - \frac{1}{2} \right)\left( n_j - \frac{1}{2} \right) + h.c. \quad \text{(SI 1)}$$

Where we have separated the short range (Hubbard) and long range parts of the interaction. The second term in eq.(SI 1) includes the disorder effects introduced by the vacancy. We diagonalize $H_0$ using the exact one particle eigenvalues:

$$H_0 = \sum_{v_i;s} \varepsilon_{v_i} c_{v_i;s}^+ c_{v_i;s} + \sum_{k;s} \varepsilon_k c_{k;s}^+ c_{k;s} \quad \text{(SI 2)}$$

The on site interaction term can be written as:

$$H_{\text{int}} = U \sum_i (n_i^2 - \vec{s}_i^2) \quad \text{(SI 3)}$$

With:

$$n_i = c_{i;\uparrow}^+ c_{i;\uparrow} + c_{i;\downarrow}^+ c_{i;\downarrow}$$

$$\vec{s}_i^2 = \frac{1}{2}(s_i^+ s_i^- + s_i^- s_i^+) + s_i^z s_i^z$$

$$s_i^+ = c_{i;\uparrow}^+ c_{i;\downarrow} \qquad\qquad \text{(SI 4)}$$

$$s_i^- = c_{i;\downarrow}^+ c_{i;\uparrow}$$

$$s_i^z = c_{i;\uparrow}^+ c_{i;\uparrow} - c_{i;\downarrow}^+ c_{i;\downarrow}$$

The local operators $c_{i;s}$ can be expressed in terms of the eigenvectors of $H_0$:

$$c_{i;s} = \sum_{v_j} \alpha_{v_j}^i c_{v_j s} + \sum_k \alpha_k^i c_{k;s}^i \quad \text{(SI 5)}$$

Interactions suppress double occupancy of the resonance. We assume that in graphite, or in graphene near the neutrality point, the resonance is half filled on the average. Then, it can be considered a spin one half interacting with the extended states. For one resonance, the Hamiltonian which describes the interaction can be written as:

$$H_{v,k} = -\frac{U}{4} \sum_{k,k'} \vec{s}_v \vec{s}_{k,k'} \sum_i \alpha_k^i \alpha_{k'}^i \left| \alpha_v^i \right|^2 - \sum_{k,k'} \vec{s}_v \vec{s}_{k,k'} \sum_{i,j} V(\vec{r}_{ij}) \alpha_k^i \alpha_{k'}^i \alpha_v^j \quad \text{(SI 6)}$$

Where the operators $\vec{s}_v$ and $\vec{s}_{k,k'}$ are generalizations of those in eq. (SI 4). The coupling is given by the exchange interactions, and it is proportional to the overlap between the wavefunctions. This Hamiltonian describes a ferromagnetic Kondo system. As the resonance is spread over many sites, $i$, the localized spin interacts with many conduction band channels, defined in terms of their angular momentum around the position of the impurity. The coupling is ferromagnetic, and the moment is not quenched at low temperatures. The effective Kondo coupling can be large, despite the low density of states of graphite near the Fermi level, because the phaseshift induced by a spin flip process in the presence can be large (S2). This enhancement of the coupling arises from the strong scalar potential induced by the vacancy. The coupling to magnetic impurities outside the graphene layers is determined by virtual transitions into states with different occupancies, leading to an antiferromagnetic coupling suppressed by the low density of states of clean graphite near the Fermi level (S3). The extension of the orbital associated to the vacancy leads to a long range coupling between magnetic moments, when the wavefunctions of resonances around different sublattices overlap. This is the case if the vacancies belong to the same sublattice. We find a ferromagnetic coupling, given by:

$$H_{v-v} = -U \vec{s}_v \vec{s}_{v'} \sum_i \left| \alpha_v^i \right|^2 \left| \alpha_{v'}^i \right|^2 - \sum_{i,j} V(\vec{r}_{ij}) \alpha_v^i \alpha_v^i \alpha_{v'}^j \alpha_{v'}^j \approx -\left( U \frac{a^2}{d^2} + \frac{e^2}{d} \right) \vec{s}_v \vec{s}_{v'} \quad \text{(SI 7)}$$

Where $d$ is the distance between vacancies, and we write for the long range part as $V(\vec{r}_{ij}) \approx e^2 / \left| \vec{r}_{ij} \right|$. The coupling between a magnetic moment and the average magnetization induced by the rest is $J \approx -U n_v a^2 - e^2 \sqrt{n_v}$ where $n_v$ is the density of vacancies. For low densities, the coupling is determined by the contribution of the long range exchange interaction. For $n_v \approx 3 \cdot 10^{11} \mathrm{cm}^{-2}$ and $e^2 \approx 1-5 \mathrm{eV}\,\overset{0}{\mathrm{A}}$, we find a Curie temperature $T_C \approx J \approx 50-250\,\mathrm{K}$. This estimate depends strongly on the density of vacancies.